\documentclass[aps,prd,twocolumn,groupedaddress,showpacs,showkeys]{revtex4-1}
\usepackage{graphicx}% Include figure files
\usepackage{dcolumn}% Align table columns on decimal point
\usepackage{bm}% bold math
\usepackage{amssymb,amsmath,latexsym,amsfonts}
\usepackage[utf8]{inputenc}
\begin{document}

%\preprint{APS/123-QED}

\title{Fluid--Gravity Correspondence under the presence of viscosity}% Force line breaks with \\

\author{B. González-Fernández}
 \email{belinkag@nucleares.unam.mx}\affiliation{Instituto de Ciencias Nucleares, Universidad Nacional Autónoma de México,\\ A. Postal
70-543, C.P. 04510, México, D.F., México.}

  % Not yet Pe\~nanietolandia

 \author{A. Camacho}
 \email{acq@xanum.uam.mx} \affiliation{Departamento de Física,
 Universidad Autónoma Metropolitana--Iztapalapa\\
 Apartado Postal 55--534, C.P. 09340, México, D.F., México.}


\date{\today}% It is always \today, today,

\begin{abstract}
The present work addresses the analogy between the speed of sound of
a viscous, barotropic, and irrotational fluid  and the equation of motion for a non--massive field in a curved manifold. It will be shown that
the presence of viscosity implies the introduction, into the equation of motion of the gravitational analogue, of a source term which
entails the flow of energy from the non--massive field to the
curvature of the spacetime manifold. The stress-energy tensor is also computed and it is found not to be constant, which is consistent with such energy interchange.
\end{abstract}

\pacs{04.50+h, 04.20.Jb, 11.25.Mj, 04.60.Bc}% PACS, the Physics and Astronomy
\maketitle

%****************************************************************************
\section{Introduction}
Analogies have always played a fundamental role in physics and
mathematics. Indeed, they define a bridge between two different
branches of science. One of the current trends in this direction is
the so--called Analogue Gravity, a topic that considers the
similarities between physics in a curved spacetime and, usually,
condensed matter systems \cite{Barc1}. The movement of light waves
in a curved manifold finds a beautiful analogy in the description of
sound waves in a moving fluid.

Besides the inherent beauty of analogies, they are also important as
they represent a powerful tool that can give us an insight of what
might be happening in a system based on phenomena that occur in its
correspondent analogue.  Of course, at this point, a warning word is
a must. Indeed, the existence of an analogy does not imply a
complete equivalence between the aforementioned theories. In other
words, the main objective of an analogy is to capture a sufficient
number of relevant features of the modeled theory.

In particular, the search for analogies between condensed matter and gravitational cases are interesting because they could
shed some light upon some major pro-blems in gravity that still remain unsolved. For instance, since condensed matter physics
is described resorting to a quantum field theory which can be tested experimentally, the analysis of
the analogies between particle physics, particularly the quantum
vacuum, and condensed matter could bring light into the physics for
trans--Planckian situations \cite{Volo1}.

One of the simplest cases involving a moving fluid and an analogue
spacetime is provided by the acoustics of the fluid \cite{Visser1}.
In this example the existence of a silent hole has been shown,
namely, a region from which sound is unable to escape. This
resembles the situation of a black hole in general relativity
\cite{Misner1}.

The analysis of the speed of sound in a fluid and the possibility of
shaping this case in the form of the equation of motion for a
non--massive field immersed in a curved spacetime is already an old
result \cite{Barc1}. The logical premises in the aforementioned
work consider a barotropic, inviscid and irrotational flow. The
result is that the equation of motion for the
corresponding velocity potential can be cast as that of a minimally--coupled scalar field, without mass, which propagates in a Lorentzian geometry. It has to be underlined
that this equation of motion has a null source term. The
gene-ralization of this result, namely, the possibility of ha-ving a
rotational fluid has also already been carried out. Here we may
state that, when non--vanishing vorticity is present in a fluid, the
analogy with gravitation can be done, but the Unruh acoustic
metric \cite{Unruh1} does not suffice, i.e.,  the introduction of
Cartan torsion is compulsory \cite{Garcia1}.

In the present work, we consider a barotropic and irrotational fluid,
but now we allow a non--vanishing visco-sity. We will show that the equation of
motion for the corresponding velocity potential can be considered as
that of a non--massive and minimally coupled field
propagating in a (3+1) Lorentzian geometry. In contrast to
\cite{Barc1}, now a source term appears which is induced by the
presence of viscosity. This last feature can be interpreted,
physically, as an element that implicates the transfer of energy from
the field to the curvature of the manifold.
\bigskip
\bigskip

%****************************************************************************
\section{Analogue Gravity and Viscosity}

In the spirit of \cite{Barc1} we now proceed to prove the following.\\

{\bf \textit{Theorem}}

Let us consider an irrotational and barotropic fluid, though with a
non--vanishing viscosity. The equation of motion for the velocity
potential associated to an acoustic perturbation, $\phi_1$, is
identical to the equation of motion of a minimally coupled
non--massive field in a $(3+1)$--Lorentzian geometry in which a
source term is present:

\begin{equation}
\partial{\mu}\Bigl[\sqrt{-g}g^{\mu\nu}\partial_{\nu}\phi_1\Bigr]=\partial{\mu}\Bigl[\sqrt{-g}g^{0\mu}\frac{\nabla^2\phi_1}{R}\Bigr].\label{ele1}
\end{equation}

The source contribution entails an energy flow from the non--massive
field to the curvature of the manifold.

{\bf \textit{Proof}}

We start writing down our equations of motion, which in our case are
the continuity equation and the Navier--Stokes equations
\cite{landau}

\begin{equation}
\frac{\partial{\rho}}{\partial t} +
\nabla\cdot(\rho\vec{v})=0,\label{ele2}
\end{equation}

\begin{equation}
\rho\frac{D\vec{v}}{Dt} =-\nabla p +\Bigl(\zeta + \frac{1}{3} \eta \Bigr)\nabla\Bigl(\nabla\cdot\vec{v}\Bigr)
+\eta\nabla^2\vec{v}.\label{ele3}
\end{equation}

Here $p$ denotes pressure (the existence of the concept of pressure
is related to the definition of ideal fluid \cite{Chorin}, $\rho$ is
the density, $\zeta$ and $\eta$ are viscosity coefficients,
$\vec{v}$ is the velocity of the fluid element, and $D/Dt=
\partial /\partial t + \vec{v}\cdot\nabla$ is the material derivative \cite{Chorin} .

We now impose the condition of vanishing vorticity

\begin{equation}
\nabla\times\vec{v}= \vec{0}.\label{ele4}
\end{equation}

Additionally, we require the fluid to be barotropic, i.e., enthalpy
has to be a function of the pressure only \cite{Barc1}. Under these
condition, and using some vector analysis formulas, we obtain that
(\ref{ele3}) reduces to

\begin{equation}
\rho\frac{D\vec{v}}{Dt} =-\nabla p +\left( \zeta + \frac{4}{3} \eta
  \right)\nabla\Bigl(\nabla\cdot\vec{v}\Bigr).\label{ele5}
\end{equation}

The barotropic behavior of the fluid means there exists a
function $h(p)$, called enthalpy, such that

\begin{equation}
h(p) =\int^p\frac{dp'}{\rho(p')},\label{ele6}
\end{equation}

hence

\begin{equation}
\frac{1}{\rho}\nabla p= \nabla h(p).\label{ele7}
\end{equation}

The final assumption involves the behavior of the coefficient
$\mu\equiv \left( \zeta + \frac{4}{3} \eta \right)$, which we assume that changes slower than
the remaining parameters.

We now define the concept of velocity potential, $\phi$, which is
well-defined due to the condition of vanishing vorticity
\cite{Chorin},

\begin{equation}
\vec{v} = -\nabla\phi.\label{ele8}
\end{equation}

Then our equation of motion can be shaped in the following form

\begin{equation}
\nabla\Bigl[\partial_t\phi - h -\frac{1}{2}(\nabla\phi)^2 +
\frac{\mu}{\rho}(\nabla^2\phi)\Bigr]=0,\label{ele9}
\end{equation}

so that it reads

\begin{equation}
\partial_t\phi - h -\frac{1}{2}(\nabla\phi)^2 +
\frac{\mu}{\rho}(\nabla^2\phi)=0.\label{ele99}
\end{equation}

We now proceed to introduce the idea of sound in the context of
fluctuations in our physical parameters around the equilibrium
state. Keeping the expansions up to li-near terms of our small
parameter, denoted here by $\alpha$, the thermodynamical variables
become, asymptotically, \cite{Holmes},

\begin{equation}
\rho= \rho_e + \alpha\rho_s, \label{ele10}
\end{equation}
\begin{equation}
p= p_e + \alpha p_s, \label{ele11}
\end{equation}
\begin{equation}
\phi= \phi_e + \alpha\phi_s, \label{ele12}
\end{equation}
\begin{equation}
h(p)= h(p_0) + \alpha\frac{p_1}{\rho_0} +... \label{ele13}
\end{equation}

Introducing these expansions into the equation of motion we obtain two
different equations: one that shows no dependence upon $\alpha$
and another that hinges linearly on $\alpha$. Indeed,

\begin{equation}
\partial_t\phi_0 - h_0 -\frac{1}{2}(\nabla\phi_0)^2 -
\frac{\mu}{\rho}(\nabla^2\phi_0)=0,\label{ele14}
\end{equation}
\begin{equation}
\partial_t\phi_1 + \vec{v}_0\cdot\nabla\phi_1 - \frac{p_1}{\rho_0} -\frac{\mu}{\rho} \nabla^2 \phi_1 =0, \label{ele15}
\end{equation}
where we have assumed $ \bigtriangledown\cdot \vec{v}_0=0 $.

In order to have a consistent system, we must introduce our
expansions in the continuity equation (see (\ref{ele2})), to obtain

\begin{equation}
\frac{\partial{\rho}_0}{\partial t} +
\nabla\cdot(\rho_0\vec{v}_0) = 0\label{ele16} ,
\end{equation}

\begin{equation}
\frac{\partial{\rho}_1}{\partial t} -
\nabla\cdot\Bigl(\rho_0\nabla\phi_1 +\phi_1\nabla\phi_0
\Bigr)=0.\label{ele17}
\end{equation}

Finally, the restriction concerning the barotropic behavior leads to

\begin{equation}
\rho_1 = \frac{\partial\rho}{\partial p}p_1.\label{ele18}
\end{equation}

Our previous results allow us to mold this last expression in the
following form

\begin{equation}
\rho_1 = \frac{\rho_0}{c^2}\Bigl[\partial_t\phi_1
+\vec{v}_0\cdot\nabla\phi_1 -
\frac{\mu}{\rho}\nabla^2\phi_1\Bigr],\label{ele19}
\end{equation}

Therefore, the  equation of motion becomes
\begin{eqnarray}
-\frac{\partial}{\partial
t}\Bigl\{\frac{\rho_0}{c^2}\Bigl[\partial_t\phi_1
+\vec{v}_0\cdot\nabla\phi_1 - \frac{\mu}{\rho_{0}}\nabla^2\phi_1\Bigr]\Bigr\}
\nonumber\\
+ \nabla\cdot\Bigl\{\rho_0\nabla\phi_1 -
\vec{v}_0\frac{\rho_0}{c^2}\Bigl[\partial_t\phi_1
\nonumber\\
+\vec{v}_0\cdot\nabla\phi_1 -
\frac{\mu}{\rho_{0}}\nabla^2\phi_1\Bigr]\Bigr\}=0.\label{ele20}
\end{eqnarray}

This equation can be cast in the following form
\begin{eqnarray}
\partial_{\mu}\left[f^{\mu\nu}\partial_{\nu}\phi_1\right] =
\partial_{\mu}\left[f^{0\mu}\frac{\mu}{\rho_{0}}\nabla^2\phi_1\right].\label{ele21}
\end{eqnarray}

In this last expression we have that
\begin{equation}
 f^{\mu\nu} \equiv \frac{\rho_0}{c^2} \left( \begin{array}{ccc}
    - 1 & \vdots & - v_0^j\\
   \cdots &  & \cdots \\
    - v_0^i & \vdots & c^2 \delta^{ij} - v_0^i v_0^j
  \end{array} \right)   . \label{fv17}
\end{equation}

In order to obtain the analogy with the case of a scalar
d'Alembertian in a curved space ($\Delta\phi =
(1/\sqrt{-g})\partial_{\mu}(\sqrt{-g}g^{\mu\nu}\partial_{\nu}\phi)$)
we deduce that our metric is provided by $f^{\mu\nu} =
\sqrt{-g}g^{\mu\nu}$, and $\sqrt{-g} = \rho^2_{0}/c$. Hence, the
equation of motion reads

\begin{eqnarray}
\frac{1}{\sqrt{-g}}\partial_{\mu}(\sqrt{-g}g^{\mu\nu}\partial_{\nu}\phi_1)=\nonumber\\
\frac{1}{\sqrt{-g}}\partial_{\mu}\Bigl[\sqrt{-g}g^{0\mu}\frac{\mu}{c^{2}}\nabla^2\phi_1\Bigr].\label{ele22}
\end{eqnarray}

\section{Stress-energy tensor}

In order to understand what is happening with energy in this system, we calculate the stress-energy tensor \cite{Misner1}.

As usual, the energy density of the fluid element will be encoded in the component $ T^{00} $, so that
\begin{equation}
T^{00}=\rho\varepsilon + \frac{1}{2} \rho v^{2}  ,\label{T00}
\end{equation}
where $ \varepsilon $ represents the internal energy per unit mass.

Meanwhile, the components $ T^{i0}=T^{0i} $, expressing the $i$-th component of the momentum, will be
\begin{equation}
T^{i0}=T^{0i}=\rho v_{i} . \label{T0i}
\end{equation}

To find the $ T^{ij} $ components of the tensor, corresponding to the flux of the $i$-th component of the momentum through a surface with  $x^{j}=constant$, we compute the rate of change of the $i$-th momentum in time
\begin{equation}
\frac{\partial}{\partial t} \left(\rho v_{i}\right)=\rho \left( \frac{\partial v_{i}}{\partial t}  \right)+\left(\frac{\partial\rho }{\partial t} \right)v_{i} . \label{tij1}
\end{equation}

Substituting the continuity equation (\ref{ele2}) and the Navier--Stokes equations (\ref{ele3}) in (\ref{tij1}), we can rewrite it as
\begin{equation}
\begin{array}{r}

\frac{\partial}{\partial t} \left(\rho v_{i}\right)=-\frac{\partial}{\partial x_{k}} \ \left\lbrace \rho v_{i} v_{k} + \left[p- \mu \frac{\partial v_{l}}{\partial x_{l}} \right] \delta_{ij} \ \right\rbrace  \\
\\
\equiv -\frac{\partial}{\partial x_{k}} \left(\Pi_{ik}\right) , \label{tij3}

\end{array}
\end{equation}
where we have conveniently defined the tensor
\begin{equation}
\Pi_{ik} \equiv \rho v_{i} v_{k} + \left[p- \mu \frac{\partial v_{l}}{\partial x_{l}} \right] \delta_{ij} .
\end{equation}
If we integrate (\ref{tij3}) over some volume $ V $,
\begin{equation}
\frac{\partial}{\partial t}\int \rho v_{i} dV = \int -\frac{\partial \Pi_{ik}}{\partial x_{k}}=
\oint \Pi_{ik} \widehat{n}_{k} dS ,\label{L7.3}
\end{equation}
it is easy to see that the left side of (\ref{L7.3}) is the rate of change of the $i$-th component of the momentum per unit time, so that $ \Pi_{ik} $ will be the flux of the $i$-th component of the momentum through a surface with  $x^{j}=constant$, thus
\begin{equation}
T^{ik}=\Pi^{ik}=\rho v_{i} v_{k} + \left[p- \mu \frac{\partial v_{l}}{\partial x_{l}} \right] \delta_{ij}.\label{Tij}
\end{equation}

Having all its components, we combine (\ref{T00}), (\ref{T0i}) and (\ref{Tij}) to form a single matrix representing the stress-energy tensor:
\begin{equation}
(T^{\mu \nu})=\left( \begin{array}{ccc}
\rho\varepsilon + \frac{1}{2} \rho v^{2} & \vdots &  \rho v_{i}\\
\cdots &  & \cdots \\
\rho v_{i} & \vdots & \rho v_{i} v_{k} + \left[p- \mu \frac{\partial v_{l}}{\partial x_{l}} \right] \delta_{ij}
\end{array} \right) . \label{Tmunu}
\end{equation}

Finally, we linearize the components of the tensor.

For $ T^{00} $, we find
\begin{equation}
T^{00}_{(0)}= \rho_0 h_{0} - p_{0} + \frac{1}{2}\rho_0 v_0^{2}  \label{T00o0}
\end{equation}
\begin{equation}
\begin{array}{r}
 T^{00}_{(1)} =\rho_1 h_{0}+  \frac{1}{2} \left(\rho_1 v_0^{2}\right) +  \frac{1}{2} \left(\rho_0 \vec{v}_0\cdot\vec{v}_1\right)  ,\label{T00o1}
\end{array}
\end{equation}
where we have expressed $ \varepsilon $ in terms of the enthalpy, using the deffinition of the latter.

In the case of $ T^{0i} $, the result is
\begin{equation}
T^{0i}_{(0)} \equiv \rho_0 v^{i}_0  \label{T0io0}
\end{equation}
\begin{equation}
 T^{0i}_{(1)} \equiv \rho_0  v^{i}_1 + \rho_1 v^{i}_0 . \label{T0io0}
\end{equation}
At last, for $ T^{ij} $ we have
\begin{equation}
T^{ij}_{(0)} \equiv \rho_0 v_0^{i} v_0^{j} +\left(p_0-\mu \nabla \cdot \vec{v}_0 \right)\delta^{ij} \label{Tijo0v}
\end{equation}
\begin{equation}
T^{ij}_{(1)} \equiv \rho_1 v_0^{i} v_0^{j} + \rho_0 \left(v_1^{i} v_0^{j} + v_0^{i} v_1^{j}\right) + \left(p_1 -\mu \nabla \cdot \vec{v}_1 \right)\delta^{ij} .\label{Tijo1v}
\end{equation}

It should be noted that, when we make $ \rho_1\rightarrow 0$,
$\vec{v}_1\rightarrow0$, and $p_{1}\rightarrow 0 $, the inviscid
case is recovered.

Harking back to our particular case, these aforementioned parameters
are not constant. In other words, clearly there is some interchange
of energy between the scalar field and the spacetime, as the first
transfers part of its energy to the second.

\section{Conclusions and discussion}

Considering the case $\zeta=0$ and $\eta=0$ we recover the usual
theorem \cite{Barc1}. Observe that, in our case, the presence of a
non--vanishing viscosity appears as a source term for the equation
of motion of the non--massive field. Let us now provide a physical
interpretation, in the context of (\ref{ele22}), for the role of
viscosity. Viscous terms provide a mechanism by which momentum
transfer becomes relevant in a fluid. Another way to understand the
role of viscosity is noting that it provides a mechanism by which
macroscopic energy can be turned into internal energy \cite{Holmes}.
This last statement allows us to comprehend, at the level of our
main result, this issue. Indeed, the aforementioned flow of energy
implies that, for our last equation, we may assert that our
parameter $\mu$, which contains the information concerning
viscosity, is responsible for the flow of energy from our field
$\phi_1$ into the curvature of the manifold, a fact already shown in
the previous section.

It is readily seen that the left hand side of our main result,
(\ref{ele22}), is associated to a non--massive and spinless field,
such as a non--massive boson with vanishing spin. In other words,
our equation would provide the dynamics of a particle with these
features moving in a curved spacetime, assuming also, of course, a
coupling with the curvature of the kind here introduced. It has to
be underlined that the flow of energy from a system into the
curvature of a manifold does not, necessarily, imply a breakdown of
the postulates of GR. For instance, in GR a system radiating
gravitational waves loses its energy, which becomes "ripples" of
spacetime, i.e., ripples of the curvature.

\begin{acknowledgements}
B. González-Fernández thanks CONACYT for the grant received.
\end{acknowledgements}

\end{document}